\documentclass[aps,prb,showpacs,twocolumn]{revtex4-1}
\usepackage{amsmath,amssymb,mathrsfs,bm,setspace}
\usepackage{graphicx}
\usepackage{hyperref}
\usepackage{color}
\newcommand{\al}{\alpha}

\newcommand{\g}{\gamma}
\newcommand{\de}{\delta}

\newcommand{\la}{\lambda}
\newcommand{\La}{\Lambda}

\newcommand{\w}{\omega}

\renewcommand{\S}{\Sigma}

\newcommand{\pd}{\partial}

\newcommand{\round}[1]{\left( #1 \right)}

\newcommand{\abs}[1]{\left| #1 \right|}
\newcommand{\cvec}[2]{\round{\begin{array}{c} #1 \\ #2 \end{array}}}

\newcommand{\beq}{\begin{equation}}
\newcommand{\eeq}{\end{equation}}
\newcommand{\Beq}{\begin{eqnarray}}
\newcommand{\Eeq}{\end{eqnarray}}
\newcommand{\bml}{\begin{multline}}

\newcommand{\eeqm}{\end{multline}}

\newcommand{\bsp}{\begin{split}}
\newcommand{\esp}{\end{split}}
\newcommand{\down}{\downarrow}
\newcommand{\up}{\uparrow}

\newcommand{\mat}[4]{\begin{pmatrix}#1&#2\\#3&#4\end{pmatrix}}
\newcommand{\eq}[2][]{\begin{equation} #2 \label{#1}\end{equation}}

\renewcommand{\b}[1]{{\bm #1}}

\newcommand{\inv}{^{-1}}

\newcommand{\mc}{\mathcal}

\newcommand{\req}[1]{Eq.~(\ref{#1})}

\newcommand{\nn}{\nonumber}

\DeclareMathOperator{\tr}{tr}
\DeclareMathOperator{\diag}{diag}
\DeclareMathOperator{\sgn}{sgn}

\begin{document}

\title{Gauge-field fluctuations in the 3D topological Mott insulator}
\author{William Witczak-Krempa$^1$, Ting Pong Choy$^1$ and Yong Baek Kim$^{1,2}$}
\affiliation{
$^1$Department of Physics, The University of Toronto, Toronto, Ontario M5S 1A7, Canada\\
$^2$School of Physics, Korea Institute for Advanced Study, Seoul 130-722, Korea}
\date{\today}

\begin{abstract}
We study the low-energy properties of three-dimensional (3D) topological Mott insulators which can be viewed as strong topological insulators of spinons interacting with a three-dimensional gauge field. The low-energy behavior of such systems is dominated by the two-dimensional (2D) gapless surface spinons coupled to the bulk gauge field. We find that a dimensional crossover from 3D to 2D in the gauge field fluctuations may occur as the system’s thickness and/or temperature is varied. In the thin sample limit, the gauge field fluctuations effectively become 2D and the problem becomes analogous to the standard 2D spinon-gauge field theory. In the 3D limit, the bulk gauge field fluctuations lead to a low-energy theory for the coupled system that is more controlled than for the pure 2D case. We discuss various experimental signatures such as the heat capacity scaling as $T \ln 1/T$ as well as modified Ruderman-Kittel-Kasuya-Yoshida interactions on the surface.
\end{abstract}

\maketitle

\section{Introduction}

Topological insulators of non-interacting fermions are characterized by topological invariants of the 
band structure.\cite{Zhang_review,Kane_review,KaneMele1,KaneMele2,bernevig1,bernevig2,Fu-Kane-Mele,Moore,Fu-Kane,Fu-Kane2,Roy1,Roy2} While the bulk spectrum is gapped, the boundary states of such systems carry the characteristic informations about the topology of the bulk band structure. This important observation has been instrumental to the experimental identifications of the 2D and 3D topological insulators.\cite{Exp0,Exp1,Exp2,Exp3,Exp4,Exp5} In particular, 3D strong topological insulators are characterized by the presence of an odd number of Dirac-cone spectra.\cite{Zhang_review,Kane_review,Fu-Kane,Fu-Kane2} These ``helical liquids" or Dirac fermions with a finite chemical potential provide low energy excitations in otherwise gapped systems. The presence of such excitations and their helical nature,
the ``locking" of momenta and spins, were beautifully confirmed in recent experiments.\cite{Exp1,Exp2,Exp3,Exp4,Exp5}

It is natural to ask what would be the analogous topological phases in interacting many-body systems. 
A well-known example of such correspondence would be the case of integer and fractional quantum Hall
states albeit they are time-reversal symmetry breaking phases in contrast to the topological insulator.\cite{dassarma}
There have been various theoretical proposals for ``interacting" topological insulators.\cite{kallin,Pesin-Balents,hur,levin,karch,mcgreevy}
Some of these theories use the so-called ``parton construction". Here the electron is ``split" into
partons carrying fractional quantum numbers, where the individual partons are in 
topological insulator phases.\cite{levin,karch,mcgreevy} 
On the other hand, in the context of Mott insulators, the construction via the 
slave-rotor formulation\cite{slave-rotor,senthil-rotor,podolsky} has been proposed,
where the electron Hilbert space is written as a product of
the spin-1/2 neutral spinon states and rotors representing charge coherence.\cite{kallin,Pesin-Balents,hur,slave-rotor,senthil-rotor,podolsky}
Because of the constraint coming from
the original electron Hilbert space, these excitations are strongly coupled to a U(1) gauge field.
The slave-rotor field theory can, for example, naturally describes an insulator-metal transition via the condensation of the rotor degrees of freedom. Moreover, the resulting Mott insulators are spin liquids where the spinons are natural low energy excitations.\cite{slave-rotor,senthil-rotor,podolsky}
When one starts from a strong topological insulator of weakly interacting electrons, instead of a metal, the increased interaction may allow the transition to a Mott insulator state by gapping out the rotor degrees of freedom (or destroying charge coherence).
The resulting state is again a spin liquid, but the spinons in this case would inherit the topological band structure of the original electrons. These ``topological" spin liquid states are called topological Mott insulators.\cite{kallin,Pesin-Balents,hur}
Recently it has been proposed that certain pyrochlore iridates materials as well as other transition metal oxides
may be good candidates for such novel phases.\cite{Pesin-Balents,shitade,BJ,vishwanath}.
One advantage of this theory is that it naturally connects the topological insulators of weakly interacting electrons 
to topological Mott insulators where electrons are no longer well-defined excitations because of strong interactions.

In analogy to the topological insulators, the boundary state should carry the characteristic information about the topology of the bulk spectra of spinons. This time, however, the helical liquid of spinons at the surface must be coupled to 3D bulk gauge field. Thus the low energy properties of topological Mott insulators are dominated by the surface helical liquid of spinons coupled to a 3D gauge field. This is in contrast to the typical gauge theory of spin liquids where the dimensionality of the spinons and gauge field are the same.

In this work, we examine the effect of singular gauge field fluctuations on the helical liquid of spinons (Dirac fermions with a finite chemical potential). As mentioned above, the gauge field fluctuations
coupled to the spinons are not confined to the surface but permeate the whole system. 
If the system is thin enough, the gauge field fluctuations associated with the direction normal to the surface are not
 excited and hence the gauge field becomes effectively two-dimensional. In which case the problem is very
analoguous to the standard spinon--gauge-boson problem believed to arise in 2D spin liquid Mott 
insulators.\cite{ybkim,ioffe,chubukov,polchinski,nayak,sslee,MS,mross}
Here the one-loop fermion self-energy $\Sigma \sim -i |\omega|^{2/3} {\rm sgn} (\omega)$ suggests that the surface spinons are not well-defined.   
On the other hand, if the 3D nature persists, we find a novel theory in which the effective 2D gauge propagator is suppressed by a square root:
$D(\nu, q)=1/\sqrt{\chi q^2+\g |\nu|/q}$ leading to a marginal renormalization of the spinons with the self-energy
correction being $\S\sim-i\w\ln (1/|\w|)$. The resulting phase is analoguous to, but not the same as, 
the $\nu=1/2$ compressible quantum Hall state with unscreened Coulomb interaction.\cite{ybkim,ioffe,nayak}

Notice that the effect of the singular gauge field fluctuations in 2D regime is known to be non-trivial even in the limit of a large number of fermion flavours ($N$): there are non-trivial structures in three and higher order loop contributions and currently the fate of the theory is not entirely known.\cite{sslee,MS,mross} Even in this case there may be a finite temperature regime where the naive one-loop or the leading order large-$N$ result may apply before the system possibly becomes unstable. 
In the 3D regime, however, the theory is better controlled as we demonstrate in a simple scaling analysis
and we expect that our results obtained here may apply even in the low energy regime.
A more detailed analysis of higher order loops may be necessary to answer this question completely, which is
beyond the scope of this work.

The low energy excitations of the topological surface spinons and the gauge fields 
lead to the heat capacity scaling as $T \ln (1/T)$ in the 3D regime. 
In contrast to a regular 3D strong topological insulator of non-interacting electrons, 
the topological Mott insulator has no charge response at the surface, for example the electrical
conductivity vanishes and there are no Friedel oscillations associated with a charge impurity. 
The gapless spinon states carry entropy, however, and would give rise to a ``metallic" thermal conductivity. 
The surface states of topological Mott insulators, just as in regular strong topological insulators, 
mediate a long-range RKKY interaction between magnetic impurities deposited on the surface.\cite{zhang_rkky,spin_helix,franz1}
The angle dependence of the generated RKKY interaction is identical in both cases because the gauge field does not break any symmetries of the helical fermion action. 
However, the scaling in terms of the distance between magnetic impurities is modified because of the gauge field
interactions: it is logarithmically suppressed from $1/r^2$ to
$1/(r^2\ln (k_Fr))$. In the 2D regime, the difference is greater and the power law 
may acquire significant corrections.\cite{ioffe}

The rest of the paper is organized as follows.
In section \ref{sec:model}, we define the low energy theory consisting of a helical liquid of spinons coupled to a gauge field.
The effective gauge field dynamics is derived in section \ref{sec:gf_dynamics}, where the crossover between the 2D and 
3D regimes is also discussed. In section \ref{sec:fermion}, the spinon self-energy is computed at the one-loop level.
In section \ref{sec:scaling}, a simple scaling analysis for the spinon--gauge-field coupling is presented. 
We compute the response functions near $2k_F$ in regular topological insulators in section \ref{sec:bare_2kf}.
This is followed by the computations of $2k_F$ response functions in topological Mott insulators in section \ref{sec:dressed_2kf},
where the resulting behavior of the RKKY interactions is discussed.
We conclude and discuss future directions in section \ref{sec:conclusion}. The appendices show details of some diagrammatic computations.
\begin{figure}
\begin{center}
\includegraphics[scale=0.6]{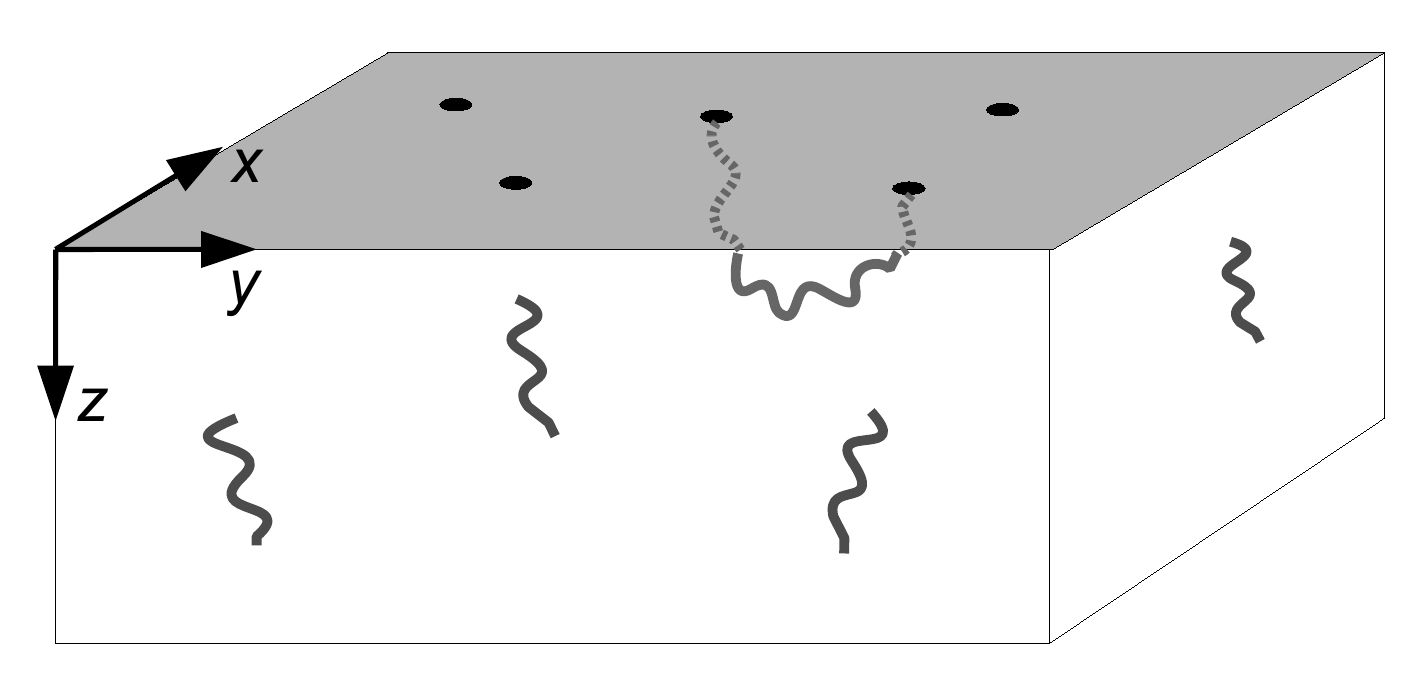}
\caption{\label{fig:system} A schematic representation of the system under consideration with its low energy excitations. 
The topologically protected gapless spinons (black dots) abide on the shaded surface at $z=0$ and are coupled to the bulk gauge
fluctutations (wiggly lines). }
\end{center}
\end{figure}
\section{Model}
\label{sec:model}
The low energy properties of the surface states in a strong topological insulator of non-interacting electrons
can be described by the Hamiltonian\cite{Zhang_review,Kane_review}
\eq[H_sti]{
H =\int d^2\b x \,\psi^\dag(\b x) [-i\b\sigma\cdot(\hat z \times\nabla) -\mu]\psi(\b x) \ ,
}
where $\b x=(x,y)$ labels inplane positions, while $z$ is the out-of-plane direction;
 $\psi$ is a 2-spinor representing the physical spin,
$\b\sigma$ is the Pauli matrix vector and $\mu$ the chemical potential relative to the Dirac point.
We use units where $\hbar=v_F=1$.
This Hamiltonian describes massless Dirac fermions at finite chemical potential, the latter breaking the emergent Lorentz invariance.

In the topological Mott insulating phase, the non-trivial band topology of the parent topological insulator state is transferred
to the spinons, which become gapless on the surface. In which case, we use \req{H_sti} as a minimal model to describe the gapless surface spinon states,
where $\psi$ carries the spin degree of freedom of the original electron but not its electric charge.
In contrast to the strong topological insulator, the surface spinons do not propagate freely but are minimally coupled to an emergent
U(1) gauge field $A_\mu(x,z)$, where we use $x=(\tau,\b x)$, $\tau$ being the Euclidean time. The action of the 
interacting spinons and gauge bosons is given by
\begin{multline}
	\mc S= \mc S_A + \int d^3x\,\psi^\dag(x)\Big\{\pd_\tau -igA_0(x,0) -\mu\\ 
	-i\b\sigma\cdot \left[\hat z \times (\nabla -ig\b A(x,0))\right]\Big\}\psi(x) \ ,
\label{tot_lagrangian}
\end{multline}
where
$S_A=(1/g_0)^2\int d^3x dz F_{\mu \nu} F_{\mu \nu}$ is the bulk Maxwell action for the 3D gauge field, with
field strength tensor $F_{\mu\nu}=\pd_\mu A_\nu-\pd_\nu A_\mu$. The coupling $g_0$ comes from the high-energy degrees of freedom
associated with the bulk fermions. We note that while the gauge bosons 
exist in the bulk, the gapless fermions abide on the surface, which we take to be at $z=0$, as can
be seen in Fig.~\ref{fig:system}. In this work, we focus on a single surface and thus ignore the time-reversal
partners located on the opposite boundary.
\subsection{Boundary conditions}
We specify the boundary conditions for the gauge field along the $z$-direction as follows
\eq[bc]{
\pd_z A_\mu(x,z)\big|_{z=0,L_z} =0, \quad \mu=0,x,y,z \ ,
}
where $L_z$ is the length of the system in the $z$-direction. This Neumann-type boundary condition 
corresponds to a ``free endpoint" type of boundary condition where the gauge field can take arbitrary 
values at the surface as long as it does not vary approaching the boundary. We note that although the
gauge field is confined to the material, we do not impose the Dirichlet-type boundary condition 
$A_\mu=0$ on the boundary. This would result in the absence of coupling between the surface fermions 
and the gauge bosons, which is not consistent with the microscopic lattice theory where the boundary fermions
are always coupled to a lattice-link variable which is a line integral of the gauge field. 

By virtue of the boundary condition, we can express the gauge field as
\eq[gf_expansion]{
A_\mu(x,z)=\sum_{n_z=0}^\infty A_{n_z,\mu}(x) \cos\left(\frac{\pi n_z z}{L_z}\right) \ ,
}
so that on the surface, we have $A_\mu(x,0)=\sum_{n_z=0}^\infty A_{n_z,\mu}(x)$.
Substituting this result into \req{tot_lagrangian}, we see that the surface fermions couple to ``standing waves" with wave numbers $n_z=0,1,2,\dots$. This additional excitation channel will play an important role in the 
low energy theory by weakening the gauge fluctuations in comparison to a purely two-dimensional theory.
\subsection{Gauge choice}
We fix the gauge as follows
\eq[gauge]{
\pd_x A_x +\pd_y A_y = 0 \ .
}
In comparison with the standard three-dimensional transverse (or Coulomb) gauge, $\pd_x A_x +\pd_y A_y+\pd_z A_z = 0$, 
we do not include the $\pd_z A_z$ term. Notice that our gauge fixing condition does not
couple $A_z$ to the remaining components. This is convenient because $A_z$ does not directly couple to 
the surface fermions. There might be an indirect coupling through the bulk fermions that are coupled to
all the components of the gauge field, but the former are gapped and such a coupling will be
irrelevant in the low energy limit. Thus $A_z$ will not play an important role in our discussion and
we shall neglect it. 
We are left with two gauge components: the temporal one ($A_0$) and the 2D-transverse one ($A_\perp$). 
We shall simply call the latter ``transverse" in the remainder of the discussion. We now turn to the renormalization
of the gauge field by the low energy particle-hole excitations of surface spinons.
\section{Gauge field dynamics}
\label{sec:gf_dynamics}
\begin{figure}
\begin{center}
\includegraphics[scale=0.52]{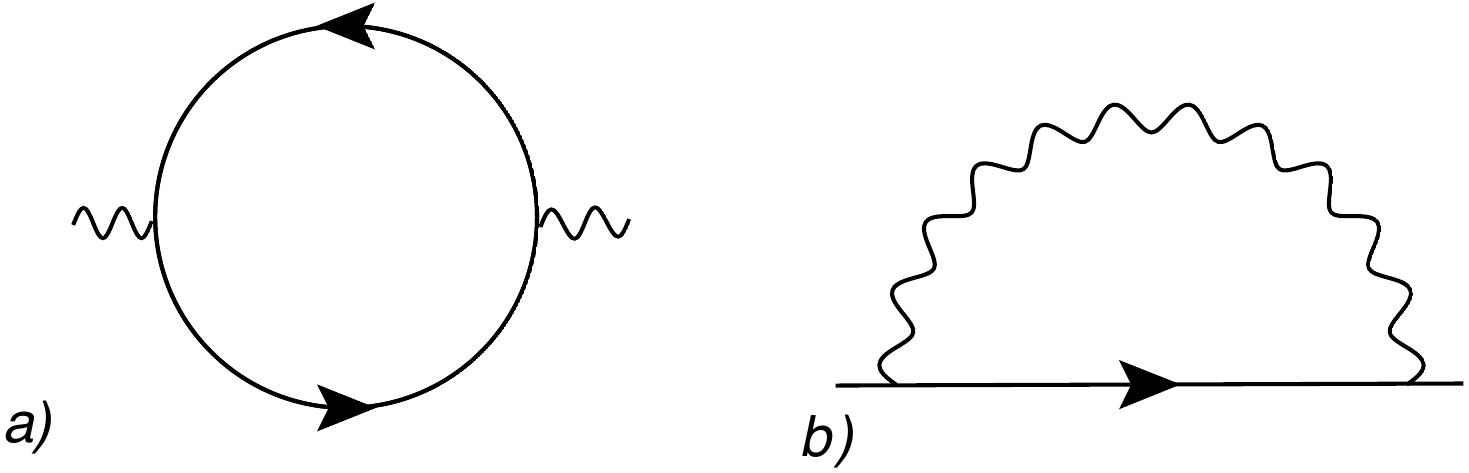}
\caption{\label{fig:self-energy} One-loop self-energy corrections to a) the gauge boson and b)
fermion. The solid lines denote the free fermions while the curly lines represent the
transverse gauge bosons.}
\end{center}
\end{figure}
We first write the free-fermion action in frequency-momentum space: $S_f=\int_p \psi_p^\dag \hat G\inv_p \psi_p$, where
we have used the energy-momentum 3-vector $p=(p_0,p_x,p_y)$
and $\int_p=\int \frac{d^3p}{(2\pi)^3}$. (It should be clear from the context when instead
we mean $p=|\b p|$.)  The inverse propagator matrix is
\begin{align}
\hat G\inv_p &=i\w +\mu +p_y \sigma_x -p_x \sigma_y \ , \nn\\
	&=\mat{i\w+\mu}{p_y+ip_x}{p_y-ip_x}{i\w+\mu} \ .
\end{align}
We can conveniently express the matrix propagator as\cite{raghu}
\eq[bare_f_propagator]{
\hat G_p=\sum_{s=\pm}\hat P_s(\theta_p)G_s(p) \ ,
}
where
\begin{align}
 \hat P_s(\theta_p)&=\frac{1}{2}\left[1+s\hat z\cdot( \hat p\times\b\sigma)\right] \ ,\nn\\
 &=\frac{1}{2}\mat{1}{-sie^{-i\theta_p}}
 		{sie^{i\theta_p}}{1} \ ; 
\end{align}
and
\eq{
G^{-1}_s(p)=i\w-\xi_{s,p} \ ,
}
with $\theta_p$ being the angle of $\b p$ relative to the $x$-axis and $\xi_{s,p}=sp-\mu$. We note that the propagator $G_-=(i\w-p-\mu)\inv$ remains finite
for $\w=0$ and $p=\mu$; it corresponds to fermions in the lower Dirac cone. Lying far from the Fermi surface, they
do not contribute to the low energy properties. We further note that if the chemical potential is tuned to zero, one is left with Dirac fermions and
the above decomposition, \req{bare_f_propagator}, is not useful; instead, one can perform the calculations by exploiting the Lorentz invariance of the action.  

We evaluate the renormalization of the gauge field components $A_0$ and $A_\perp$ due to the gapless surface spinons
by calculating the bosonic self-energy at one loop (Fig.~\ref{fig:self-energy}(a)).
It will be useful to consider a general polarization function with incoming and outgoing
interaction vertices $\hat\g_{a}$ and $\hat\g_{b}$, respectively. The
vertices are matrices in spin space and can in general dependent on momentum. 
The corresponding polarization function reads 
\begin{align}
\Pi_{ab}(q)&=\int_p\tr[ \hat\g_a\hat G_p\hat\g_b\hat G_{p+q}]\ , \\
&=g_ag_b\sum_{s,s'=\pm}\int_p G_s(p)G_{s'}(p+q)f_{ab}^{ss'} \ ,
\label{pol_ftn}
\end{align}
where $g_{a,b}$ are the coupling strengths and
\begin{equation}
f_{ab}^{ss'} = \tr[ \hat\g_a\hat P_s(\theta_p) \hat\g_b\hat P_{s'}(\theta_{p+q})]/g_ag_b \ ,
\label{form_factor}
\end{equation}
is a form factor arising from the combination of the angle dependence of
the fermionic propagators and the vertices. It will in general be different for various polarization functions,
as we shall see. It has also been normalized so as to be independent of the coupling strengths.

We consider first the boson self-energy for the transverse gauge component, $A_\perp$. One can simply see that the vertices
are momentum dependent: $\hat\g_{a(b)}=\pm g\hat q\cdot\b \sigma$. 
The expression for the polarization function becomes
\eq[pi_11]{
\Pi_{\perp\perp}(q)= g^2\sum_{s,s'=\pm}\int_p G_s(p)G_{s'}(p+q)f_{\perp\perp}^{ss'} \ ,
}
where 
\eq{
2f_{\perp\perp}^{ss'}= -1+ss'\frac{p\cos 2\theta+q\cos\theta}{|\b p+\b q|},
}
and $\theta=\hat p\cdot\hat q$ is the angular integration variable. In the limit where
the external frequency-momentum vector is zero, we find $\Pi_{\perp\perp}(q_\mu=0)=-g^2\Lambda$,
where $\Lambda$ is an arbitrary ultraviolet (UV) cutoff. We regularize this expression so that the 
transverse component remains massless, as is required by gauge invariance \cite{ybkim,ioffe}. For finite $q_\mu$,
we obtain
\beq
	-\Pi_{\perp\perp}= \chi q^2+\g\frac{|\nu|}{q} \ ,
\eeq
where $\chi=c_0g^2/\mu$ and $\g=c_1g^2\mu$; the $c_i$ are positive real numbers. This result is valid for small momenta $q\ll k_F$ and in the static limit $|\nu|\ll q$.

Next, we turn to the polarization function for the temporal component $A_0$. The vertices are simply $\hat\g_{a(b)}=g\hat 1$ because
$A_0$ couples to the spinon density; as a result the form factor becomes
\eq[form_factor_00]{
2f_{00}^{ss'}=1+ss'\frac{p+q\cos\theta}{|\b p+\b q|}.
}
We find
$\Pi_{00}(q_\mu= 0)\propto-g^2\mu$, i.e. the Fermi surface of spinons induces a mass for the temporal gauge component. 
We note that this mass does not correspond to a UV cutoff,
and cannot be removed by the above regularization. We also note that the mixed polarization
functions vanish exactly $\Pi_{0\perp}=\Pi_{\perp 0}=0$. At finite chemical potential, we can thus neglect
the massive (screened) temporal gauge component in comparison with the transverse one, which
mediates a long-range interaction between the spinons. 

We now write down the effective gaussian action (or RPA action) for the transverse 
gauge field by retaining the dominant contributions from the bulk Maxwell Lagrangian and from the 
self-energy due to the helical fermions:
\begin{multline}
	\mc S_{A_\perp,\rm eff}=\sum_{n_z}\int_q \abs{A_{n_z,\perp}(q)}^2 \\
	\times\left[ \chi_z\left(\frac{\pi n_z}{L_z}\right)^2 + \chi q^2+\gamma \frac{|\nu|}{q}\right],
\end{multline}
where $\chi_z=L_z/g_0^2$. The gauge propagator is then
\eq[3d_gauge_propagator]{
\mc D_{n_z}(q)=\frac{1}{\chi_z\left(\frac{\pi n_z}{L_z}\right)^2 + \chi q^2+\gamma \frac{|\nu|}{q}} \ .
}
The key difference from the finite chemical potential QED$_3$ is the appearance of $n_z$ encoding the
possible fluctuations of the gauge field in the out-of-plane direction. We note that in the effective action we have
only retained terms with identical incoming and outgoing standing wave numbers as these processes 
dominate in the low-energy limit.
\subsection{2D-3D crossover and effective gauge field propagator}
When we evaluate amplitudes for quantum processes involving virtual gauge field fluctuations such as the fermion self-energy, 
the fermion--gauge-boson vertex, etc, we will need to sum over
all possible bulk standing waves of the transverse gauge-field. Since the fermions have 
no dependence on the standing wave number we can simply sum the gauge field propagator over all $n_z$ to obtain 
an \emph{effective 2D} propagator:
\eq{
	D(q)=\sum_{n_z}\mathcal D_{n_z}(q) \ .
}	
In the limit of infinite $L_z$, the sum becomes an integral; it can be evaluated
simply by rescaling the integration variable:
\begin{align}
D(q)&=\frac{L_z}{\pi}\int_0^\infty dq_z \frac{1}{\chi_zq_z^2+\mathcal D_0(q)\inv} \ ,\\
&=\frac{g_0L_z^{1/2}}{2}\sqrt{\mathcal D_0(q)} \ ,
\end{align}
where $\mathcal D_0(q)\inv=\chi q^2 +\g|\nu|/q$ is the standard, inverse gauge field propagator
in the absence of bulk fluctuations. Note that the effective 2D gauge field propagator is less singular
as $q_\mu\rightarrow 0$ than the purely 2D propagator $\mathcal D_0(q)$. This is to be expected
as the flucutations have access to more phase space.

For finite $L_z$, the discrete sum can be done exactly: $D(q)=h\left(g_0^2L_z\mathcal D_0(q)\inv\right)$,
with $h(x)=(1+\sqrt x\coth\sqrt x)/2x$.
The function $h$ interpolates between the purely two- and three-dimensional regimes
\eq{
h(x) = \frac{1}{2}\begin{cases}
  2/x & \text{, $x\ll 1$, \quad\quad 2D regime} \\
  1/\sqrt x & \text{, $x\gg 1$, \quad\quad 3D regime} \ .
  \end{cases}
}
Let us now examine the crossover between these two regimes. First, we consider the zero
temperature case. The condition $x\ll 1$ for the occurence of the 2D regime corresponds to
\eq{
\chi q^2+\g\frac{|\nu|}{q}\ll \frac{1}{g_0^2L_z} \ .
}
In the limit of infinite inplane area, one can always satisfy this inequality provided
one takes $|q_\mu|$ sufficiently small, i.e. for small enough $|q_\mu|$ 
the fluctuations behave in a purely 2D fashion, as is expected since the 
inplane dimensions are infinite while the out-of-plane one is finite. For a finite system, it would be more meaningful to put a lower
bound on the inplane momenta, $q_i>2\pi/L$, where $L$ characterizes the size of the inplane dimensions. We can further use the scaling relation $\nu\sim (\chi/\g) q^3$ for the Landau damped gauge field. Combining everything we obtain $L_z\ll \mu L^2/g^2g_0^2$. 

At finite temperature, one can use $\nu\sim T$ together with $\nu\sim (\chi/\g) q^3$ to obtain the following inequality
for the occurence of the 2D regime for the gauge field fluctuations
\eq{
L_z\ll \frac{1}{g^2g_0^2\mu^{1/3} T^{2/3}} \ .
}
As expected, as one increases the temperature, the sample would have to be thinner in order for the gauge
fluctuations to be of two-dimensional nature.

To summarize, the gauge fluctuations can be of either 2D nature (strong) or 3D nature (weak) depending on the system parameters $L_z$, $T$, $\mu$. The effective 2D propagator corresponding to each of these regimes is
\begin{align}
	D(q)=\begin{cases}
	\dfrac{1}{\chi q^2+\g |\nu|/q} & \text{,\quad\quad 2D regime}\ , \\[.35cm]
    \dfrac{d_0}{\sqrt{\chi q^2+\g |\nu|/q}} & \text{,\quad\quad 3D regime} \ ,
	\end{cases}
\end{align}
with $d_0=g_0L_z^{1/2}/2$. As can be noted, in the 2D regime the low energy theory for the topological Mott insulator becomes analoguous to
the standard spinon--gauge-field problem, which, as mentionned in the introduction, was extensively studied and still lacks
a satisfactory understanding. On the other hand, the 3D regime has weaker gauge fluctuations and this will make the theory
more tractable. In the remainder of the work, we shall focus mainly on the 3D regime.
\subsection{Heat capacity from gauge fluctuations}
Using the gaussian action for the transverse component of the gauge field, the low temperature heat capacity can be shown to scale like $C\sim T \ln 1/T$ in the 3D regime while it scales like $C\sim T^{2/3}$ in the 2D regime.\cite{ybkim_heatCap} In the absence of the topological surface states, the main contribution to the heat capacity would come from the gapless gauge bosons, yielding a much smaller contribution scaling like $T^3$. The phonons would also contribute the same power law.
\section{Spinon self-energy}
\label{sec:fermion}
We evaluate the one-loop spinon self-energy due to gauge field fluctuations.
As is well known, the main effect of the over-damped gauge field is the appearance of 
non-analytic frequency dependence of the self-energy. The angular dependence of the fermionic
propagator, however, is not changed as the coupling to the gauge field does not break the symmetries
of the original free fermion system (this is discussed in detail later in this section). 
Formally, the one-loop self-energy matrix reads (Fig.~\ref{fig:self-energy}(b))
\begin{align}
\hat\Sigma(p)&= g^2\int_q D(q)(\hat q\cdot\b\sigma)\hat G(p+q)(\hat q\cdot\b\sigma) \ , \label{fermion_self_energy}
\end{align}
where the fermion propagators are given by \req{bare_f_propagator}. We quickly note that
$G_{\up\up}=G_{\down\down}$ so that $\Sigma_{\up\up}=\Sigma_{\down\down}$. The off-diagonal elements
are also simply related: $\Sigma_{\up\down}(\w,\b p)^*=\Sigma_{\down\up}(-\w,\b p)$.
For $\w\sim 0$ and $p\sim\mu$, we find (see appendix \ref{app:fermion_self_energy}) 
\beq
\hat\Sigma(p) = \mat{1}{-ie^{-i\theta_p}}{ie^{i\theta_p}}{1}\Sigma_{\up\up}(\w),
\eeq
with the same frequency dependence for all components of the self-energy matrix:
\eq{
\Sigma_{\up\up}(\omega)=-i\begin{cases}
\la_2 \sgn(\w)|\w|^{2/3}, &\text{2D} \\
\la_3 \w \ln\tfrac{\mu}{|\w|}, &\text{3D} \ .
\end{cases}
}
$\la_{2,3}$ are positive constants: $\la_2 =c_2\mu^{1/3}$ and
$\la_3=c_3g\mu^{1/2}g_0L_z^{1/2}$, 
where the $c_i$ are positive order one numbers. We see that in the 2D limit, where the 
gauge fluctuations are the strongest, we obtain the standard scaling $\sim|\w|^{2/3}$.\cite{ybkim,ioffe}
In the 3D limit, we instead obtain the marginal correction $\sim\w\ln|\w|$, which is only logarithmically larger
than the bare dynamical term. 

The Dyson equation with one-loop self-energy, $\hat{\mathcal G}\inv=\hat G^{-1} - \hat\Sigma$, yields the dressed fermionic propagator 
\eq[dressed_f_propag]{
\hat{\mc G}_p = \sum_{s=\pm} \hat P_s(\theta_p) \mc G_s(p) \ ,
}
where the dressed function $\mc G_+$ reads
\eq[dressed_scalar_propag]{
\mc G_+\inv=i\w -\Sigma_{\up\up}(\w) -\xi_p \ ,
}
while $\mc G_-=G_-$ remains unchanged at leading order. As expected the angle-dependence of the dressed propagator
remains the same as for the bare one, while the dynamical term is renormalized.

The above result is consistent with symmetries. Indeed, the free fermion Hamiltonian, \req{H_sti},
is invariant under continuous rotations of the momentum and spin of the fermions by the same angle. From a more general field theory perspective, this spin-momentum rotation is a Lorentz transformation, under which the spinor naturally transforms under a two-dimensional representation of the symmetry group.
Consider such a clockwise rotation by an angle $\al$:
\begin{align}
	\b x &\rightarrow R_\al\b x; \nn\\
	\psi(x_0,\b x) &\rightarrow e^{i\frac{\al}{2}\sigma^z}\psi(x_0,R_\al\b x) \ , \\
	\psi^\dag(x_0,\b x) &\rightarrow \psi^\dag(x_0,R_\al\b x)e^{-i\frac{\al}{2}\sigma^z} \ ,
\end{align}
where $R_\al=e^{i\al\tau^y}$, $\tau^\mu$ denoting Pauli matrices acting in coordinate space; 
$e^{i\frac{\al}{2}\sigma^z}=\diag(e^{i\al/2},e^{-i\al/2})$. The action
depends only on the relative angle between the spin and momentum of fermions, via the term $\hat z\cdot(\sigma\times\nabla)$, and thus
 remains invariant under a joint rotation of the spin and momentum by the same angle $\al$. The same holds true in the presence
 of the gauge field, the components of the latter transforming in the same way as the coordinates. The
   symmetry requires $\langle e^{i\frac{\al}{2}\sigma^z}\psi_p \psi_p^\dag e^{-i\frac{\al}{2}\sigma^z}\rangle=\langle \psi_{p_0,R_\al \b p} \psi_{p_0,R_\al \b p}^\dag \rangle$. The left hand side can be shown to give:
\eq{
	 e^{i\frac{\al}{2}\sigma^z}\langle\psi_p \psi_p^\dag\rangle e^{-i\frac{\al}{2}\sigma^z} =\mat{\mathcal G_{\up\up}}{e^{i\al}\mathcal G_{\up\down}}{e^{-i\al}\mathcal G_{\down\up}}{\mathcal G_{\down\down}}_p\ ,
}
which agrees with the angle dependence obtained for the fermionic propagators as can be seen by letting $\theta_p\rightarrow \theta_p-\al$ in \req{dressed_f_propag}. 
\section{Scaling analysis}
\label{sec:scaling}
In this section, we provide a scaling analysis of the spinon--gauge-boson system in
the 3D regime in order to analyze the effect of the gauge fluctuations.
We start with the free fermionic action which can be written as 
\eq{
S_f=\int_p \psi^\dag_p[i\w +\mu+p_y\sigma^x-p_x\sigma^y]\psi_p \ .
}
We carry our analysis on a patch of the Fermi surface near $(k_F,0)$;
we express the momenta as deviations from this point: $\b p=(k_F+k_x,k_y)$, where $|k_x|,|k_y|\ll k_F$. The analysis
is further simplified in a basis that diagonalizes the free-fermion action:
\eq[S_f_newBasis]{
S_f=\int_p f_p^\dag\mat{i\w-\xi_p}{0}{0}{i\w+\mu+p}f_p \ ,
}
where the new 2-spinor, $f^\top=(f_+,f_-)$, is related to the original one by a SU(2) transformation
\eq{
\cvec{\psi_\up}{\psi_\down}=\frac{1}{\sqrt{2}}\mat{1}{ie^{-i\theta_p}}{ie^{i\theta_p}}{1}\cvec{f_+}{f_-} \ .
}
We recognize that the diagonal elements of the new action, \req{S_f_newBasis}, correspond to the inverse propagation amplitudes for fermions in the upper and lower Dirac cones, $G_\pm\inv$. At finite chemical potential (and temperature less than $\mu$), we can omit the $f_-$ fermions as they are gapped. The entire action expanded near $(k_F,0)$ reads
\begin{multline}
	S=\int d^3k\; \bar f_k(i\nu-v_xk_x-l_yk_y^2)f_k \\
	+ \int dk_z d^3k\; \left(\g \frac{|\nu|}{|k_y|}+\chi k_y^2+\chi_z k_z^2\right)|A_{k_z}(k)|^2\\
	+g\int dq_zd^3qd^3k\; A_{q_z}(q)\bar f_{k+q} f_k \ ,
\end{multline}
where, at the bare level, $v_x=1$ and $l_y=1/2\mu$. We have omitted the subscript $+$ from the fermion fields. $A_{k_z}(k)$ is the $x$-component of the gauge field, i.e. the one that couples the most with the fermions at 
$(k_F,0)$. Note that we are using the continuous limit for bulk excitations of the gauge field.

We first rescale the frequency and momenta:
\begin{align}
\nu'&=b\nu \ ,  \\
k_x'&=bk_x \ ,  \\
k_y'&=b^{1/z}k_y \ ,  \\
k_z'&=b^{1/z}k_z  \ .
\end{align}
We note that not a single dynamical exponent $z$ will keep the entire action invariant because the fermions have a dynamical exponent $z_f=2$ (with respect to $k_y$) while the bosons have $z_b=3$. 

If we make the choice $z=3$, we find
\begin{align}
[l_y]&=1/3\ , \\
[g]&=0\ ,
\end{align}
with all other couplings being marginal. 

If instead we use $z=2$ all couplings in the fermionic action naturally remain marginal. One then has a choice for the scaling
exponent of the gauge field by either requiring that the $k_y^2$ term or the Landau damping term remain invariant. 
The first choice would be the natural one had we started with the minimal action for the gauge field, i.e. the same
action as above except with the omission of the Landau damping term (which is generated in 
the one-loop bosonic self-energy).
Thus keeping the $k_y^2$ term invariant, we obtain the following engineering dimensions 
\begin{align}
[l_y]&=0\ ,  \\
[\g]&=1/2\ ,  \\
[g]&=0\ ,
\end{align}
with all other couplings being marginal. 
Notice that both scaling choices ($z=3$ and $z=2$) make the coupling to the gauge field
marginal, which is consistent with the one-loop self-energy correction acquiring only a logarithmic correction.
With these naive scaling results in mind, 
we now study the RG flow at the one-loop level. 
Following Mross \emph{et al.}\cite{mross}, we integrate out the modes whose momentum lies in the shell $\La/b^{1/z}<|k_y|<\La$,
$\La\sim k_F$ being the UV cutoff.
To order $g^2$, the fermionic dynamical term gets modified by the self-energy diagram to 
\eq{
	i\nu(1+cg^2\ln b)\rightarrow i\nu b^{cg^2} \ ,
}
where $c= 1/(\pi^2\chi^{1/2}z)$. Note that this leading correction is independent of the Landau damping of the gauge boson
so that it is the same whether one includes the Landau damping term or not.
The correction to the the vertex vanishes because at zero external frequency the integrand of the $k_x$ integral has its two poles in the same half-plane. The dynamical term of the gauge field remains invariant because the integration of high-$|k_y|$ modes 
leads to the generation of irrelevant operators.

The self-energy correction can be incorporated in a modified frequency scaling
\eq{
\nu'=\nu b^{1+cg^2} \ .
}
We use $z=3$ and keep the same scaling for $k_x, k_y$ and $k_z$ as above. Using $b=e^l$, the differential RG
equations read
\begin{align}
	{dl_y\over dl}&={l_y\over 3}\ , \\
	{d\g\over dl}&=-cg^2\g\ , \\
	{dg^2\over dl}&=-cg^4\ .
\end{align}
We see that the coupling $g$ has become marginally irrelevant. 
If instead we use the $z=2$ scaling, we obtain
\begin{align}
	{d\g\over dl}&=\frac{\g}{2}-cg^2\g\ , \\
	{dg^2\over dl}&=-cg^4\ ,
\end{align}
with all other parameters remaining invariant.
Thus, in both cases, the coupling to the gauge field is marginally irrelevant and the theory seems
well controlled once the dressed fermion propagator with the one-loop self-energy is used. 

Now that we have a better understanding of the low-energy properties of the coupled system, we turn to the investigation
of the $2k_F$ polarization functions, which are related to the physically relevant RKKY interactions on the surface of the TMI.
\section{Bare $2k_F$ correlation functions for the surface
states of a topological insulator}
\label{sec:bare_2kf}
We give the results for the bare, static, spin-spin polarization functions near $2k_F$, whose fourier transform
yields the spinon mediated RKKY interactions between surface impurities. We note that if we neglect the gauge fluctuations, the polarization functions are naturally the same as for a strong topological insulator, as given by\cite{franz1} for instance. More precisely, we will evaluate \req{pol_ftn}, with vertices $\sigma^i$, $i=x,y,z$. We evaluate the form factors, defined in \req{form_factor}, in the limit appropriate for
 the external momentum being near a nesting vector of the Fermi surface, $q\sim 2k_F$. Retaining only $s=s'=+$ and defining $f_{ij}\equiv f_{ij}^{++}$, with $i,j\in \{x,y,z\}$, to leading order we find that these form factors do not depend on the 
internal angle but only on the external one, $\theta_q$. We can thus write 
\eq{
\Pi_{ij}(0,\b q) = f_{ij}(\theta_q)\,\Pi(0,q)\ ,
}
where
\eq{
	\Pi(0,q)=\int_p G_+(p)G_+(p+q)\ ,
}
and the angle dependence reads
\beq
f_{ij}=\begin{pmatrix}\cos^2\theta_q &\tfrac{1}{2}\sin{2\theta_q} & -i\cos\theta_q\\
\tfrac{1}{2}\sin{2\theta_q} & \sin^2\theta_q &-i\sin\theta_q \\
i\cos\theta_q & i\sin\theta_q & 1 \end{pmatrix} \ . 
\label{pol_angle_dep}
\eeq
The non-analytic dependence near $2k_F$ obeys the standard scaling in two dimensions with exponent $1/2$:
\eq{
\Pi(0,q)\sim \mu\begin{cases}
1, & \de_q<0 \\
1 - \sqrt{\de_q/k_F}, & \de_q>0 \ ,
\end{cases}
}
where $\de_q=q-2k_F$. This result is valid for $|\de_q|/k_F\ll 1$. 
One can perform a Fourier transform to real space
\eq{
\chi_{ij}(\b r) =\int d^2 \b q \ \Pi_{ij}(0,\b q) e^{i\b q\cdot\b r} \ .
}
This leads to a RKKY interaction term between two impurity spins $\b S_{1(2)}$ located at $\b r_{1(2)}$ of the following form:
\begin{multline}
	H_{\rm RKKY}=J_1(r) S_1^zS_2^z+J_1(r)(\b S_1\cdot\hat r)(\b S_2\cdot\hat r)\\
	+J_2(r) \hat r\cdot (\b S_1\times \b S_2)\ ,
\end{multline}
where $\b r=\b r_1-\b r_2$. For a sufficiently large separation between the impurity spins, $2k_F r\gg 1$, we have
\begin{align}
	J_1(r) &\propto \frac{\sin 2k_Fr}{r^2} \nn , \\
	J_2(r) &\propto \frac{\cos 2k_Fr}{r^2} \ .
\end{align}
We have dropped subleading terms decaying as $1/r^3$. These induced RKKY
interactions on the surface of a strong topological insulator were also examined by\cite{zhang_rkky,spin_helix}
using a different approach.
\section{Dressed $2k_F$ correlation functions for the topological Mott insulator}
\label{sec:dressed_2kf}
In this section we will determine the effects of the gauge fluctuations on the RKKY interactions on the surface of a topological Mott insulator and
we will show that they differ from the regular topological insulator. 
\subsection{Vertex corrections}
\begin{figure}
\begin{center}
\includegraphics[scale=0.7]{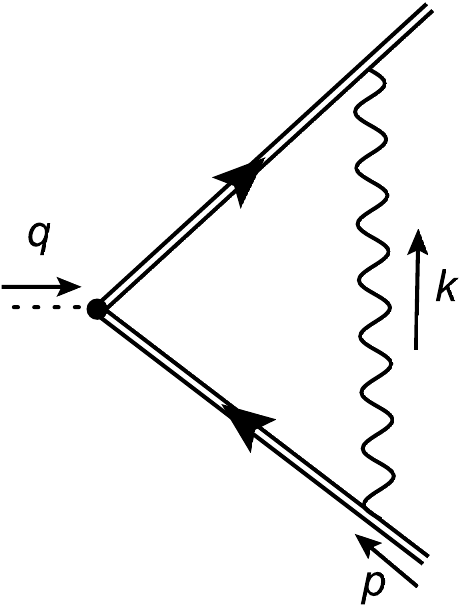}
\caption{\label{fig:vertex} $2k_F$ vertex correction. The black dot represents any spin vertex and we use
 double lines to represent dressed fermionic propagators. }
\end{center}
\end{figure}
We first examine the effect of low-energy gauge bosons on the $2k_F$ vertices. 
For a general vertex $\hat\g_a$, the correction we need to evaluate at one loop is (Fig.~\ref{fig:vertex}):
\begin{multline}
\de\hat\g_a(q,p)=-\int_k \hat\g_\perp(\hat k) \hat{\mc G}(p+q-k)\hat\g_a \\
\times\hat{\mc G}(p-k)\hat\g_\perp(-\hat k)D(k) \ .
\end{multline}
We use the dressed fermionic propagators.
We set the external frequency $q_0$ to zero. The singular
behavior is expected to occur when the incoming fermion is on the Fermi surface, i.e. $p\sim k_F$, and is
backscattered by the external boson so that $\b q\sim -2\b p$. The internal transverse gauge fluctuation
has small $k_\mu$. As before, the vertex involving the latter is simply $\hat\g_\perp(\hat k)=g\hat k\cdot \b\sigma$.
Using the decomposition for the fermionic propagators we can write it more simply as
\begin{multline}
\de\hat\g_a=-g^2\sum_{s,s'}\int_k\mc G_s(p-k)\mc G_{s'}(p+q-k)
 D(k)\hat F_a^{ss'}\, ,
\end{multline}
where $\hat F_a^{ss'}$ is a matrix form factor: 
\eq{
	\hat F_a^{ss'} = (\hat k\cdot\b\sigma)\hat P_{s'}(\theta_{p+q-k})\hat\g_a \hat P_s(\theta_{p-k})(-\hat k\cdot\b\sigma) \ .
}
Note that the form factor is a polynomial of cosines and sines with constant coefficients. Although it can never diverge, it can 
potentially suppress the amplitude. In appendix \ref{app_vertex}, we show that the one-loop vertex correction with the form factor matrix set to a constant is not singular in the infrared (IR) when $q_0=0,q=2k_F,p_0=0,\b p=-\b q/2$. Thus the vertex corrections $\de\hat\g_i$, $i=x,y,z$, will not alter the non-analycity near $2k_F$.
This results from the suppression of the effective gauge propagator in the 3D regime and not from the smearing of quasiparticles. Indeed, the vertex correction computed with bare fermionic propagators is also IR-convergent.

\subsection{$2k_F$ spin-spin correlators and RKKY interaction}
\begin{figure}
\begin{center}
\includegraphics[scale=0.45]{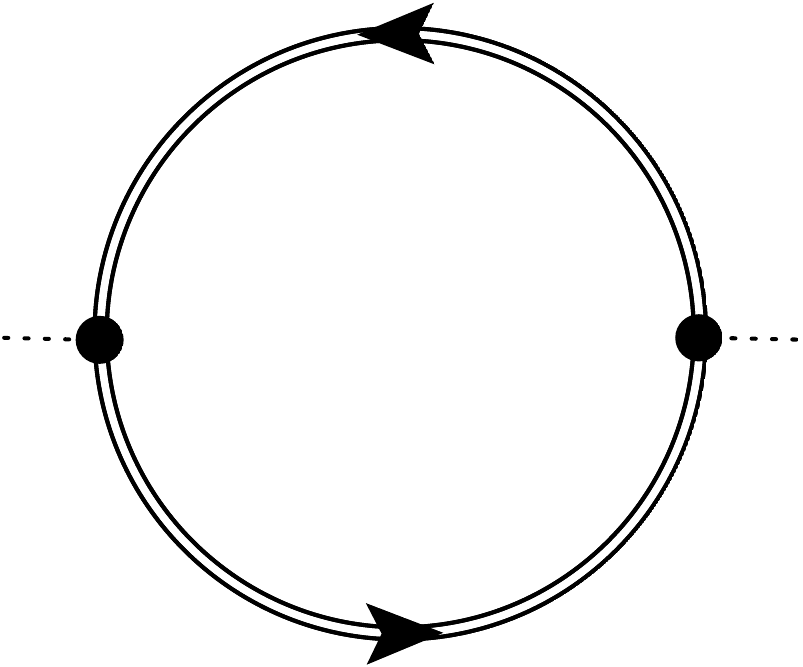}
\caption{\label{fig:dressed_pol} $2k_F$ dressed polarization function. The double line represents the dressed fermionic propagator. }
\end{center}
\end{figure}
The gauge fluctuations will affect the $2k_F$ correlation functions due to the smearing of the quasiparticles. The result will 
be a suppression of the non-analyticities. The vertex corrections, lacking singularities, will not compensate this effect. The angle dependence they carry, see \req{2kf_vertices}, will not result in new angle dependence for the 
$2k_F$ correlators as we demonstrate in appendix \ref{app:angle_dep_dr_pol}. This is in agreement with the fact that the gauge field does not break 
any symmetries present in the free fermionic action, hence the spin-spin correlation functions should retain the same angle dependence
after the inclusion of gauge fluctuations.

The renormalized polarization functions in the TMI read (Fig.~\ref{fig:dressed_pol})
\eq{
\Pi_{ij}^{\rm r}(q)=\int_p \tr [\hat\g_i\hat{\mc G}_p \hat\g_j\hat{\mc G}_{p+q}] \ ,
} 
where we have used the dressed fermionic propagators. As we argued above, the angle dependence of the 
later is the same as in the free theory, which results in the same form factors $f_{ij}$, \req{pol_angle_dep}. Thus,
$\Pi^{\rm r}_{ij}(0,\b q) = f_{ij}(\theta_q)\,\Pi^{\rm r}(0,q)$. 
Next, we need to evaluate the two-point correlator with form factor set to one:
\begin{align}
\Pi^{\rm r}(0,q)&= \int_p \mc G_+(p)\mc G_+(p+q) \ , \nn\\
	&\propto-\mu^{1/2}\int_0^\mu d\w \;\Im\; \frac{1}{\sqrt{\tfrac{1}{2}\de_q+i\lambda_3 \w\ln\frac{\mu}{\w}}}\ ,
\end{align}
where we have evaluated the momentum integrals and dropped the bare dynamical term.
The remaining integral cannot be done analytically, however one can extract the leading singular momentum dependence:
\eq{
\Pi^{\rm r}_{\rm sing}(0,q)\propto \frac{|\de_q|^{1/2}}{\ln\frac{\mu}{|\de_q|}} \ .
}
This can be seen to lead to a logarithmic modification of the real space scaling found at the bare level. Indeed, the singulariy at $2k_F$ is logarithmically weakened and this will lead to a real space scaling that decays faster. Instead of the bare scaling $1/r^2$, we get $1/(r^2\ln k_Fr)$.

In the 2D regime, we find that at one-loop the spin vertices are logarithmically enhanced\cite{ioffe}: $\de\g\propto\ln\mu/|\de_q|$,
where $\de\g$ is the $2k_F$ vertex without form factor, see \req{delta_gamma}.
In this case, the vertices are singular and will contribute to the spin-spin correlation
functions. One can resum the logarithms arising from all the ladder diagrams to obtain a power law.\cite{ioffe}
The exact exponent, however, depends on the pre-factor of the logarithm and its value has been known only
in some limits.\cite{ioffe} This would also modify the exponent of the RKKY interaction. For the density-density correlation, 
the form factor can be shown to render the vertex IR convergent.
\section{Conclusion and discussion}
\label{sec:conclusion}
We considered the effect of gauge field fluctuations in topological Mott insulators, which can be
regarded as spin liquids with the bulk spinons possessing a topologically non-trivial band structure. 
The manifestation of the non-trivial topology of the bulk-spinon spectra is the presence of
the gapless surface-spinon state, a helical liquid of spinons.\cite{Pesin-Balents}
The low energy properties of this system, therefore, are dominated by the surface spinon
state coupled to a three-dimensional emergent U(1) gauge field in the bulk.

Notice that this is a unique situation: in the usual spin liquid state derived from spin models,
the dimensionality of the gauge field is the same as that of the spinons.\cite{ybkim,ioffe}
In the finite thickness geometry, the three-dimensional nature of the gauge field is represented
by the ``standing waves" associated with the degrees of freedom perpendicular to the surface.
It is shown that there would be a crossover between the 2D and 3D regimes in the behavior of 
the gauge field, depending 
on the thickness of the system and temperature.

In the 3D regime, where the fluctuations in the bulk are strong, the effective 2D gauge field propagator can be obtained after summing over all standing-wave modes.
 It turns out that this ``holographic" 2D propagator $D(\nu, q)=1/\sqrt{\chi q^2+\g |\nu|/q}$
is less singular than its 2D counterpart $D_0(\nu, q)=1/(\chi q^2+\g |\nu|/q)$.
As a result, the spinon self-energy correction 
due to the gauge field is only logarithmically singular, $\Sigma \sim -i \omega \ln (1/|\omega|)$.
It was shown that this leads to a more controlled theory for the spinon--gauge-field system, analogous to
the case of the $\nu=1/2$ compressible state in the quantum Hall regime.
In the 2D regime, however, the one-loop self-energy $\Sigma \sim -i |\omega|^{2/3} {\rm sgn} (\omega)$
suggests that the surface spinons are not well-defined. Moreover, according to recent developments
in the theory of 2D spinon--gauge-field system, the ultimate fate of the theory is not known even 
in the large-$N$ limit.\cite{sslee,MS,mross} The same problem exists in the 2D regime in our case and we do not
attempt address this question in our current work. We emphasize, however, that the 3D regime
is better controlled as we argued in the main part of the paper.

It is shown that the 3D regime leads to $T \ln (1/T)$ specific heat behavior and the RKKY
interaction between magnetic impurities on the surface would be suppressed by a logarithmic factor. 
Perhaps the most fundamental difference
between the surface state of the usual topological insulators and topological 
Mott insulators is that both the Friedel oscillations of charge density and the RKKY
interaction exist on the surface of the topological insulator while only the latter
(with logarithmic corrections) arises in the topological Mott insulator.
It is also pointed out that the ``metallic" thermal transport should exist at the surface
of the topological Mott insulator due to the entropy carried by the surface spinons
while there is no charge transport.

One important topic that we have not discussed so far is the presence of the $\theta$ term in
the effective gauge field action.\cite{moore_nature,witten,sczhang_mono,vanderbilt,franz2} 
Because of the topological band structure of the spinons,
the so-called axion term, ${\theta \over 2\pi} \b E \cdot \b B$, would arise in the effective
action of the emergent gauge field, where the action is $2 \pi$ periodic in $\theta$.
This means that the monopoles in the compact U(1) gauge theory would possess both 
``electric" and ``magnetic" charges of the emergent gauge field.\cite{witten} 
The condensation of these
dyons will drive the confinement-deconfinement transition.
In the deconfined phase of the topological Mott insulator, these dyons are gapped
and do not play an important role in the low energy dynamics.
Thus the low energy behaviors discussed in our work should not be changed.
In the confined phase, however, the dyon condensation may lead to novel 
symmetry-breaking ground state (of usual electrons) due to mixed quantum numbers
carried by them, which would be an interesting topic for future study. 

\section*{Acknowledgements}
We thank Sung-Sik Lee for helpful discussions.
This research was supported by NSERC of Canada, the Canada Research Chair program, 
and the Canadian Institute for Advanced Research.
\appendix
\section{Fermion self-energy}
\label{app:fermion_self_energy}
We evaluate the fermionic self-energy, \req{fermion_self_energy}:
\begin{align}
\hat\S &\approx g^2\int_q D(q) G_+(p+q)\mat{0}{e^{-i\theta_q}}{e^{i\theta_q}}{0} \nn\\
&\qquad\times \mat{1}{-ie^{-i\theta_{p+q}}}{ie^{i\theta_{p+q}}}{1}\mat{0}{e^{-i\theta_q}}{e^{i\theta_q}}{0} \ , \\
&=g^2\int_q D(q)G_+(p+q)\mat{1}{ie^{i\theta_{p+q}-i2\theta_q}}{-ie^{-i\theta_{p+q}+i2\theta_q}}{1} \ .
\end{align}
To leading order $\theta_{p+q}\approx\theta_p$ since $p\sim k_F$ and $q\sim 0$. Also the relative angle between the $\b p$ and $\b q$, $\theta_p-\theta_q$, is near $\pm\pi/2$, so that the scattered fermion stays near the Fermi surface. This leads to
\eq[s-e_integral]{
\hat\S=g^2\mat{1}{-ie^{-i\theta_p}}{ie^{i\theta_p}}{1}\int_qD(q)G_+(p+q) \ .
}
We can see that the off-diagonal elements have the same angle dependence as the bare inverse fermion propagator, which leads to no new angle
dependence in the dressed propagator. Although we have used approximations to determine the angle dependence of the self-energy, it should be emphasized that
this holds exactly, as we have verified explicitly. It also follows from symmetry considerations (see section \ref{sec:fermion}).

 The remaining integral is given by $-i\lambda_3 \w\ln\mu/|\w|$. We show the main steps of the calculation. Given the external fermionic momentum vector $\b p$, we decompose the bosonic internal momentum $\b q$ along directions parallel and perpendicular to it: $q_\parallel=\b q\cdot\hat p$ and $q_\perp=\b q\cdot(\hat z\times\hat p)$. For $p\sim \mu$ and $q\ll\mu$, we have $\xi_{p+q}\approx q_\parallel+\frac{q_\perp^2}{2\mu}$. We drop $q_\parallel$ in the bosonic propagator: 
$D(q)= d_0/\sqrt{\chi q_\perp^2+\g|\nu/q_\perp|}$.
The remaining integral in \req{s-e_integral} is given by 
\begin{align}
	&\int_{-\infty}^\infty \frac{d\nu}{2\pi}\frac{dq_\parallel}{2\pi}\frac{dq_\perp}{2\pi} \frac{d_0}{\sqrt{\g\frac{|\nu|}{|q_\perp|}+\chi q_\perp^2}} \frac{1}{i(\w+\nu)-(q_\parallel+\frac{q_\perp^2}{2\mu})} \ , \nn\\
	&\sim -i \frac{d_0}{\chi^{1/2}}\int\frac{d\nu}{2\pi}\sgn(\w+\nu)\ln\frac{\mu}{|\nu|} \ ,\nn\\
	&\propto -i \w\ln\frac{\mu}{|\w|} \ ,
\end{align}
where we have performed first the $q_\parallel$ integral by contour integration (using a regulator, for instance), then extracted the leading logarithm from the $q_\perp$ integral. 
\section{Vertex corrections near $2k_F$}
\label{app_vertex}
We show the calculation of the vertex corrections. We illustrate the analysis for $\de\hat\g_z$.
 At leading order we have
\begin{multline}
\de\hat\g_z=-g^2\int_k \mc G_+(p-k)\mc G_+(p+q-k)D(k)\hat F_z^{++} \ ,
\end{multline}
where the form factor is
\begin{align*}
	\hat F_z^{++} &= (\hat k\cdot\b\sigma)\hat P_+(\theta_{p+q-k})\sigma^z \hat P_+(\theta_{p-k})(-\hat k\cdot\b\sigma)\\
	&\approx-\frac{1}{4}\mat{0}{e^{-i\theta_k}}{e^{i\theta_k}}{0} \mat{1}{-ie^{-i\theta_{-p}}}{ie^{i\theta_{-p}}}{1}\mat{1}{0}{0}{-1}\\
	&\qquad \times\mat{1}{-ie^{-i\theta_{p}}}{ie^{i\theta_{p}}}{1} \mat{0}{e^{-i\theta_k}}{e^{i\theta_k}}{0}\ ,\\
	&\approx\frac{1}{2}\mat{1}{ie^{-i\theta_q}}{ie^{i\theta_q}}{-1}  \ .
\end{align*}
In going from the first to the second equality, we have dropped $k$ as it will give rise to subleading corrections. We 
are also setting $\b q=-2\b p$. In the last equality we have used $\theta_p-\theta_k\approx \pm\pi/2$, i.e. the fermion couples most strongly
to a gauge fluctuation whose momentum is perpendicular to its own. We see that to leading order the form factor matrix is independent
of the internal momentum, hence there can be no suppression for this vertex. It remains to calculate the remaining integral with form factor matrix omitted. For $q=2k_F$, the vertex will receive its largest contribution when the fermions lie near the Fermi surface,
so that the fermionic momenta are near $\pm\b q/2$, which are antipodal points of the Fermi surface. We expand the fermionic dispersion
relations about these two points: $\xi_{\pm\frac{q}{2}-k}=\pm k_\parallel+k_\perp^2/2\mu$, where $k_\parallel=-k_F\hat q\cdot\b k$ and $k_\perp$ is the momentum in the transverse direction. We evaluate the vertex correction using $p_0=0$ and $\b p=-\b q/2=-k_F\hat q$. We drop the bare fermionic dynamical term and retain only the self-energy correction, which reads $\Sigma=-i\la_3\nu\ln\mu/|\nu|$.
The details of the calculations are
\begin{align}
	\de\g &\equiv-g^2\int_k\mc G_+(p-k)\mc G_+(p+q-k)D(k)\ ,\label{delta_gamma}\\
	&=-g^2\int_k \frac{1}{-\S+k_\parallel-k_\perp^2/2\mu} \frac{1}{-\S-k_\parallel-k_\perp^2/2\mu} \nn\\
	&\qquad\qquad\times\frac{d_0}{\sqrt{\chi k_\perp^2+\g|\nu/k_\perp|}}\ , \nn\\
	&\sim -g^2\int_0^\infty d\nu dk_\perp 
	\frac{d_0}{\sqrt{\chi k_\perp^2+\g|\nu/k_\perp|}}\frac{i\S}{(i\S)^2+(k_\perp^2/2\mu)^2}\nn \ .
\end{align}
By performing the $k_\perp$ integral first and then the frequency integral it can be seen that there is no IR divergence. This is contrast to the
half-filled fractional quantum hall scenario, where the propagator is $D\inv=\chi|k_\perp|+\g|\nu/k_\perp|$. In that case one obtains, using dressed fermions, a weak IR divergence $\sim (\ln|\ln \La_{\rm IR}|)^2$.

We deduce that the all the one-loop vertex corrections are IR convergent near $2k_F$. We conclude by giving the explicit angle dependence for the 
three spin-vertex corrections:
\begin{align}
	\de\hat\g_z &= \frac{1}{2}\mat{1}{ie^{-i\theta_q}}{ie^{i\theta_q}}{-1}\de\g \ ,\nn\\
	\de\hat\g_+ &= \frac{1}{4}\mat{-ie^{i\theta_q}}{1}{e^{i2\theta_q}}{ie^{i\theta_q}}\de\g\ , \nn\\
	\de\hat\g_- &= \frac{1}{4}\mat{-ie^{-i\theta_q}}{e^{-i2\theta_q}}{1}{ie^{-i\theta_q}}\de\g\ ,
	\label{2kf_vertices}
\end{align}
where $\de\g>0$ is a constant. Note that $\de\hat\g_\pm$ corresponds to the bare vertex $\sigma^\pm=(\sigma^x\pm i\sigma^y)/2$.
\section{Angle dependence of $2k_F$ polarization functions with dressed vertices}
\label{app:angle_dep_dr_pol}
We explicitly verify that the angle dependence obtained for the vertex corrections in the previous appendix
does not lead to a different angle dependence in the polarization functions near $2k_F$ when compared with the
free-fermion theory, \req{pol_angle_dep}. We shall only verify this for the $\Pi_{zz}^{\rm r}$ polarization function, the
calculation being very similar for the other polarization functions. Including the vertex corrections, the polarization
function reads
\eq{
\Pi_{zz}^{\rm r}(q)=\int_p \tr [\hat\g_z(p,q)\hat{\mc G}_p \hat\g_z(-p,-q)\hat{\mc G}_{p+q}] \ ,
} 
where $\hat\g_z(p,q)=\sigma^z+\de\hat\g_z(p,q)$. We evaluate the polarization function at zero external
frequency and for the external momentum equal to a nesting vector of the Fermi surface: $q_0=0$ and $q=2k_F$.
This amplitude will receive its largest contribution when the particle and hole lie near the Fermi surface,
so that the fermionic momenta are near $\pm\b q/2$. In the previous section, we have shown that the
one-loop vertex correction is not singular near $2k_F$, hence, to leading order, we can evaluate it for $p_0=0$ and $\b p=-\b q/2=-k_F\hat q$, as given in \req{2kf_vertices}. We are thus left with 
\begin{multline}
\Pi^{\rm r}_{zz}(0,\b q)=\int_p\mc G_+(p)\mc G_+(p+q) \\
\times\tr [\hat\g_z(\hat q)\hat P_+(\theta_p) \hat\g_z(-\hat q)\hat P_+(\theta_{p+q})] \ ,
\end{multline}
where we have again retained only the term with $s=s'=+$ and have defined $\hat\g_z(\hat q)=\hat\g_z(p_0=0,\b p=-\b q/2;q_0=0,\b q=2k_F\hat q)$. To obtain the angle dependence we evaluate the trace in the last equation. To leading order we can take $\theta_p=\theta_{-q/2}$ and $\theta_{p+q}\approx\theta_{q/2}$. In which case, the trace becomes independent of the integration
variable and one can evaluate it to find that it is a constant independent of $\theta_q$, just as in the bare case. This is consistent 
with the fact that the gauge field does not break any symmetry present in the free fermion action. 


\end{document}